\documentstyle[seceq,mbf,preprint]{ptptex}
 \notypesetlogo 
 \def\ind{\indent}
 \def\nn{\nonumber}
 \def\be{\begin{equation}}
 \def\ee{\end{equation}}
  \def\beq{\begin{eqnarray}}
 \def\eeq{\end{eqnarray}}
 \def\hx{{\hat x}}
 \def\hL{{\hat {\mbox {${\cal L}$}}}}
 \def\hS{{\hat S}}
 \def\hp{{\hat p}}
 \def\btheta{{\bar \theta}}
 \def\htheta{{\hat \theta}}
 \def\hvarphi{{\hat \varphi}}

 \def\bea*{\begin{eqnarray*}}
 \def\eea*{\end{eqnarray*}}
 \def\ba{\begin{array}}
 \def\ea{\end{array}}
 
 \def\cL{{\mbox {${\cal L}$}}}
 \def\cC{{\mbox {${\cal C}$}}}
 \def\cP{{\mbox {${\cal P}$}}}
 \def\cT{{\mbox {${\cal T}$}}}
 \def\plank{\hbar}
 
\title{%
Discrete Symmetries In Lorentz-Invariant Non-Commutative QED
}
\vspace{5mm}
\author{%
   Katsusada {\sc Morita}
}
\inst{%
{\it  Department of Physics, 
  Nagoya University, Nagoya 464-8602, Japan}
}
\vspace{5mm}
\abst{%
It is pointed out that the usual $\theta$-algebra
assumed for non-commuting coordinates
is not $P$- and $T$-invariant,
unless one {\it formally} transforms the
non-commutativity parameter $\theta^{\mu\nu}$
in an appropriate way.
On the other hand, the Lorentz-covariant DFR algebra,
which `relativitizes' the $\theta$-algebra
by replacing $\theta^{\mu\nu}$
with a second-rank antisymmetric tensor operator $\htheta^{\mu\nu}$,
is $\,P$- and $T$-invariant.
It is then proved that
$C, \,P$ and $T$ are separately conserved in 
Lorentz-invariant Non-Commutative QED.
}
\begin{document}
\maketitle
\section{Introduction}
Recently, there have been a lot of
works\cite{1)} on the non-commutative quantum field theory (NCQFT),
simply called QFT$_*$ in this paper.
Their aims are vast in both philosophy and methodology, 
from string theory connection
as initiated by Seiberg and Witten\cite{2)} to
phenomenological search of Lorentz violation\cite{3)}
inherent in QFT$_*$.
It is, therefore, difficult to
put them together
in a single catchword, but
we consider it worthwhile to
pursue the program along this line of thought,
which may shed light on the divergence difficulty
through modification\cite{4),5),6),7),8),9)} of the notion
of the space-time structure underlying QFT
in spite of the fact that
the modification is conceptually radical,
putting it as a `top-down' theory which
leaves us many challenges ahead, like Lorentz invariance\cite{10),11),12),13)}, 
general covariance\cite{14),15)}, 
unitarity\cite{16)}, causality\cite{17)}, analyticity\cite{18)}, 
$CPT$ theorem\cite{19),20),21)}, 
spin-statistics relation\cite{20),21)}, asymptotic 
conditions and so on.
\\
\ind
QFT$_*$ is a QFT on the non-commutative space-time
characterized by the $\theta$-algebra,
\be
[\hx^\mu,\hx^\nu]=i\theta^{\mu\nu}
\label{eqn:1-1}
\ee
where $\hx^\mu\;(\mu=0,1,2,3)$ are the hermitian
coordinate operators and
$(\theta^{\mu\nu})$ is a real antisymmetric 
constant matrix. 
Any field in QFT$_*$ becomes an operator,
$\hvarphi(\hx)$.
In terms of 
the Weyl symbol $\varphi(x)$ defined through
\beq
\hvarphi(\hx)&=&\displaystyle{1\over (2\pi)^4}\int\!d^4kd^4x\varphi(x)
e^{-ikx}e^{ik{\hat x}},
\label{eqn:1-2}
\eeq
QFT$_*$ becomes
a nonlocal
field theory on the ordinary space-time with
the point-wise multiplication of the field variables
being replaced by the Moyal $\ast$-product,
\begin{eqnarray}
\displaystyle{1\over (2\pi)^4}
\int\!d^4ke^{ikx}{\rm tr}[\varphi_1(\hx)\varphi_2(\hx)
e^{-ik{\hat x}}]&=&\varphi_1(x)*\varphi_2(x)\nn\\[2mm]
&=&e^{\frac i2
\partial_1\wedge \partial_2}\varphi_1(x_1)
\varphi_2(x_2){\big|}_{x_1=x_2=x},
\label{eqn:1-3}
\eeq
with $\partial_1\wedge \partial_2=
\partial_{1\mu}\theta^{\mu\nu}\partial_{2\mu}$
and the normalization tr$e^{ik{\hat x}}=(2\pi)^4\delta^4(k)$.
Thus 
the action defining QFT$_*$ is given by
\be
S=\int\!d^4x\cL(\varphi(x), \partial_\mu\varphi(x))_*,
\label{eqn:1-4}
\ee
where the subscript of the Lagrangian indicates that
the $\ast$-product should be taken for all products
of the field variables.
\\
\ind
Needless to say,
{\it the symmetry of
the action} (\ref{eqn:1-4}) {\it should also be the symmetry
of the $\theta$-algebra} (\ref{eqn:1-1}) {\it
if it is the symmetry of the theory and, conversely,
if a transformation leaves the action} (\ref{eqn:1-4})
{\it invariant but is not a
symmetry transformation of the $\theta$-algebra,
it is not the symmetry of the theory.} The internal
symmetry is independent of the $\theta$-algebra.
This is no longer the case for the external transformations.
In particular,
the Lorentz symmetry, one of the most fundamental symmetries in
(relativistic) QFT, is violated in QFT$_*$. To be more specific,
since the $\theta$-algebra is not Lorentz-covariant,
it is possible to take the matrix $(\theta^{\mu\nu})$ 
in a canonical form
\beq
(\theta^{\mu\nu})=\left(
\ba{cccc}
0&\theta_{\rm e}&0&0\\
-\theta_{\rm e}&0&0&0\\
0&0&0&\theta_{\rm m}\\
0&0&-\theta_{\rm m}&0\\
\ea
\right).
\label{eqn:1-5}
\eeq
The symmetry group of the $\theta$-algebra (\ref{eqn:1-1})
is then $O(1,1)\times SO(2)\bowtie T_4$, \cite{21)} where
$\bowtie$ denotes the semi-direct product.
This is only a subgroup of the Poincar\'e group
and thus makes it impossible to classify the asymptotic states
in terms of the unitary irreducible representations
of the Poincar\'e group. 
This implies, for instance, that
the tachyonic states to be
excluded from
the asymptotic states by the spectral condition
in QFT
may be classified into `massive' states
according to the symmetry group
$O(1,1)\times SO(2)\bowtie T_4$,
which appear in the intermediate states of a closure relation.\cite{18),21)}
Moreover,
Lorentz-covariant fields can not be defined except for the
Lorentz scalar field which belongs to the one-dimensional representation
of the Lorentz group. We also note that
there are two length parameters in (\ref{eqn:1-4}),
both of which should be put zero to recover the commutative limit
where Lorentz invariance holds true and the renormalization
program works. 
It is natural, however, to suppose from the correspondence principle
point of view that
there exists only one length parameter
which goes to zero in the commutative limit.
In fact, 
Snyder\cite{4)} showed that there exists a Lorentz-invariant
NC space-time in which there is a Lorentz scalar fundamental length
whose zero-limit reduces the NC space-time to the continuum one.
\\
\ind
Existence of a single Lorentz scalar fundamental length $a$
is incorporated into the above scheme if
we assume that the non-commutativity
parameter $\theta^{\mu\nu}$ is {\it not} constant
but a second-rank antisymmetric tensor.
For we can then simply put
\beq
\theta^{\mu\nu}=a^2\btheta^{\mu\nu},
\label{eqn:1-6}
\eeq
where $a$ has a dimension of length
\footnote{Our introduction of the fundamental length
is kinematical in comparison with a dynamical
meaning of the NC
scale of the order of the Planck length.\cite{6)}
There is a long history about the dynamical
aspects of the minimum length.\cite{22)}}
and
$\btheta^{\mu\nu}$ is a dimensionless,
second-rank antisymmetric tensor.
The commutative limit is attained by letting $a\to 0$.
One can then no longer put
the parameters $\theta_{\rm e}$ and
$\theta_{\rm m}$ in (\ref{eqn:1-4})
zero independently
to recover the commutative limit. This may raise
a difficulty concerning the Hamiltonian formalism
of QFT$_*$,
but we remind the readers that
the fundamental length, if any,
is reconciled with relativity only if
the notion of the continuous time-development
is abandoned.\cite{23)}
\\
\ind
It is impossible, however, to
regard $\theta^{\mu\nu}$ in the $\theta$-algebra (\ref{eqn:1-1})
as a $c$-number tensor if $\hx^\mu$ transforms as a 4-vector.
To see this 
let $U(\Lambda)$ be the unitary operator
of the Lorentz transformation
$$
\hx'{}^\mu=
U(\Lambda)\hx^\mu U^{-1}(\Lambda)
=\Lambda^\mu_{\;\;\nu}\hx^\nu.
$$
Sandwiching both sides of (\ref{eqn:1-1})
by the unitary operator $U(\Lambda)$ and its inverse
we have for a $c$-number $\theta^{\mu\nu}$
$$
[\hx'{}^\mu,\hx'{}^\nu]=
\Lambda^\mu_{\;\;\rho}\Lambda^\nu_{\;\;\sigma}
[\hx^\rho,\hx^\sigma]
=\Lambda^\mu_{\;\;\rho}\Lambda^\nu_{\;\;\sigma}
i\theta^{\rho\sigma}=i\theta^{\mu\nu}.
$$
This equation holds true only if $\theta^{\mu\nu}=0$
for $\Lambda^\mu_{\;\;\nu}=\delta^\mu_{\;\;\nu}+\omega^\mu_{\;\;\nu},
\omega_{\mu\nu}=-\omega_{\nu\mu}$ being
not identically vanishing.
This reflects a well-known fact of there
being no constant antisymmetric second-rank tensor.
Consequently, we must look for another NC algebra
which naturally provides us with a tensor $\theta^{\mu\nu}$
and preserves Lorentz invariance of the theory.
Such a Lorentz-invariant QFT$_*$ 
was formulated by
Doplicher, Fredenhagen and Roberts (DFR)\cite{6)}
motivated by the space-time uncertainty relation,
and rediscovered by Carlson, Carone and Zobin\cite{11)}
in a search to avoid
Lorentz violation in QFT$_*$.
It is based on the DFR algebra,
\beq
[{\hat x}^\mu,{\hat x}^\nu]&=&i{\hat\theta}^{\mu\nu},\;\;\;
[{\hat\theta}^{\mu\nu},{\hat x}^\nu]=0=
[{\hat\theta}^{\mu\nu},{\hat\theta}^{\rho\sigma}],\;\;\;
\quad\mu,\nu,\rho,\sigma=0,1,2,3.
\label{eqn:1-7}
\eeq
Here $\htheta^{\mu\nu}$ is a second-rank antisymmetric
tensor operator. Feynman rules of the theory
are formulated by Filk\cite{7)}
in an irreducible representation
of the DFR algebra.
It should be noted\cite{13)} that
the tensor property of the operator $\htheta^{\mu\nu}$
is proved by assuming an extra
piece in the Lorentz generators, 
which is to be added to the ordinary
orbital angular momentum part,\cite{4)}
and
the $\theta$-algebra (\ref{eqn:1-1})
holds true as a `weak' relation,
\beq
\langle\theta|[\hx^\mu,\hx^\nu]|\theta\rangle
=i\theta^{\mu\nu}\langle\theta|\theta\rangle
\label{eqn:1-8}
\eeq
where $\theta^{\mu\nu}$ is an eigenvalue
of the operator $\htheta^{\mu\nu}$.
Thus $\theta^{\mu\nu}$ is a second-rank antisymmetric
tensor. This makes it possible to define the Lorentz scalar
$a$ through (\ref{eqn:1-6}).\cite{12)}
\\
\ind
Lorentz-invariant action of QFT$_*$
\be
\hS
=\int\!d^4xd^6\theta
W(\theta)\cL(\varphi(x,\theta), \partial_\mu\varphi(x,\theta))_*,
\label{eqn:1-9}
\ee
is a simple revision of (\ref{eqn:1-4})
as shown in Ref. 11).
Here $\varphi(x,\theta)$ is the Weyl symbol
of the operator $\hvarphi(\hx,\htheta)$
defined on the DFR algebra (\ref{eqn:1-7}) and
$W(\theta)$ is a Lorentz-invariant,
normalized weight function.
The Moyal $*$-product corresponding to the operator
product $\varphi_1(\hx,\htheta)\varphi_2(\hx,\htheta)$
is given by
\begin{eqnarray}
&&\displaystyle{1\over (2\pi)^{10}}
\int\!d^4kd^6\sigma e^{ikx+i\sigma\theta}
{\rm tr}[\varphi_1(\hx,\htheta)
\varphi_2(\hx,\htheta)
e^{-ik{\hat x}-i\sigma\htheta}]\nn\\[2mm]
&&\quad\quad=
W(\theta)\varphi_1(x,\theta)*\varphi_2(x,\theta)
=W(\theta)e^{\frac i2
\partial_1\wedge \partial_2}\varphi_1(x_1,\theta)
\varphi_2(x_2,\theta){\big|}_{x_1=x_2=x},
\label{eqn:1-10}
\eeq
where $\sigma\theta\equiv \sigma_{\mu\nu}\theta^{\mu\nu}$
and we have normalized tr$e^{i\sigma\htheta}=
{\tilde W}(\sigma)$, which is the
Fourier component of $W(\theta)$,
$$
W(\theta)=\frac 1{(2\pi)^6}
\int\!d^6\sigma {\tilde W}(\sigma)e^{-i\sigma\theta}.
$$ 
Feynman rules that take into account
all irreducible representations
are yet to be formulated,
though a model calculation was attempted
in Ref. 12), where the invariant damping factor
instead of the oscillating Moyal phase was found in a 
NC scalar model.
\\
\ind
The purpose of the present paper
is limited to investigate
the discrete symmetries based on the DFR algebra (\ref{eqn:1-7}),
since the previous considerations\cite{19),20),21)} on the subject
are all based on the $\theta$-algebra (\ref{eqn:1-1}).
The next section
discusses the 
discrete transformations of the $\theta$-algebra.
The discrete symmetries in QED$_*$
are reinvestigated in the section 3 following Ref. 19).
The discrete symmetries of the DFR algebra
are proved in the section 4.
Based on this proof
the discrete symmetries in a Lorentz-invariant version\cite{12)} of QED$_*$
are discussed in the section 5.
The last section involves
short comments.
\section{Discrete transformations of the $\theta$-algebra}
Let us define parity $P$ and time-reversal $T$ 
on the operator coordinates by
\begin{eqnarray}
\cP\hx^\mu \cP^{-1}&=&\hx_\mu\;\;\;{\rm and}\;\;\;
\cT\hx^\mu \cT^{-1}=-\hx_\mu,
\label{eqn:2-1}
\end{eqnarray}
respectively, where $\hx_\mu=g_{\mu\nu}\hx^\nu$
with $(g_{\mu\nu})=(+1,-1,-1,-1)$.
Thanks to the Weyl transform (\ref{eqn:1-2})
this in fact induces the transformation,
$x\to x_P\equiv (x^\mu)_P=x_\mu$
and $x\to x_T\equiv (x^\mu)_T=-x_\mu$
under $P$ and $T$, respectively.
Under $P, T, C$ we have
\be
\cP[\hx^\mu, \hx^\nu]\cP^{-1}
=[\hx_\mu, \hx_\nu],\;\;\;
\cT[\hx^\mu, \hx^\nu]\cT^{-1}
=[-\hx_\mu, -\hx_\nu],\;\;\;
\cC[\hx^\mu, \hx^\nu]\cC^{-1}
=[\hx^\mu, \hx^\nu].
\label{eqn:2-2}
\ee
Since the $c$-number $\theta^{\mu\nu}$
pass through the operators, $\cP, \cT, \cC$,
like the unitary operator of the
Lorentz transformations,
the $\theta$-algebra
(\ref{eqn:1-1}) do not respect $P$ and $T$ but is $C$-invariant.
Nonetheless we may
assume the following
transformations,\cite{24)}
\begin{eqnarray}
\!\!\!\!\!\!\!\!\!
\theta^{\mu\nu}\to \theta_{\mu\nu}\;({\rm under}\;\;P),\;
\theta^{\mu\nu}\to -\theta_{\mu\nu}\;({\rm under}\;\;T),\;
\theta^{\mu\nu}\to \theta^{\mu\nu}\;({\rm under}\;\;C),
\label{eqn:2-3}
\end{eqnarray}
which imply
\begin{eqnarray}
\theta^{\mu\nu}\to -\theta^{\mu\nu}\;\;\;({\rm under}\;\;CPT),
\label{eqn:2-4}
\end{eqnarray}
to {\it formally} recover the discrete symmetries of the $\theta$-algebra.
Remember that time reversal is anti-unitary.
It should be noted, however,
that the transformations (\ref{eqn:2-3}) and
(\ref{eqn:2-4}) except for a trivial case $C$
can not be derived from the $\theta$-algebra
itself. 
Assuming (\ref{eqn:2-3}) and
(\ref{eqn:2-4}) in the $\theta$-algebra
amounts to assuming $\theta^{\mu\nu}$
to be a $c$-number tensor also under the improper
Lorentz transformations
(with extra minus sign under $T$), 
while we have seen that
it is impossible to regard it as a $c$-number 
Lorentz tensor.
\\
\ind
In passing we remark that
the canonical
commutation relations are invariant
under the translations, the spatial rotations and the discrete
transformations, $P, T$. In particular,
we do not need change
the sign of the Planck constant under $T$
because time reversal transformation is anti-unitary. 
If time reversal transformation were instead assumed to be 
unitary for the present purpose only,
we would have to change
the sign of the Planck constant,
$\plank\to -\plank$ under $T$.
This is possible only
if we fudge up 
Planck `operator' 
\bea*
[\hx,\hp]=i{\hat\plank}
\eea*
with $\cT{\hat\plank}\cT^{-1}=-{\hat\plank}$.
Let $|\plank\rangle$ be an eigenstate of the operator
${\hat\plank}$ with the eigenvalue $\plank$.
Since $\cT{\hat\plank}|\plank\rangle
=\cT{\hat\plank}\cT^{-1}\cT|\plank\rangle
=-{\hat\plank}\cT|\plank\rangle=\cT\plank|\plank\rangle=\plank\cT|\plank\rangle$,
$\cT|\plank\rangle$ is the eigenstate
with the eigenvalue $-\plank$, resulting in the required sign 
change of the Planck constant under $T$ (assumed to be unitary) .
Promoting $\theta^{\mu\nu}$ to
an operator $\htheta^{\mu\nu}$
does a similar job.
\\
\ind
In the next section we accept (\ref{eqn:2-3}) and
(\ref{eqn:2-4}) to prove the discrete symmetries
of QED$_*$.
It is {\it necessary}, however,
to change the sign 
of the NC parameter,
\begin{eqnarray}
\theta^{\mu\nu}\to -\theta^{\mu\nu}\;\;\;{\rm under}\;C
\label{eqn:2-5}
\end{eqnarray}
to prove $C$ and $CPT$ invariance of QED$_*$ action.
This was observed by Sheikh-Jabbari.\cite{19)}
We shall use the sign change
(\ref{eqn:2-5}) also in the relativistic version.
\section{Discrete symmetries in QED$_*$}
First of all it would be instructive to 
reinvestigate\cite{19)} the discrete symmetries in QED$_*$ action.
\\
\ind
The NC extension of the free Dirac action
is given by
\beq
S_{D_{0*}}&=&
\int\!d^4x{\bar\psi}(x)*
(i\gamma^\mu\partial_\mu-m)\psi(x).
\label{eqn:3-1}
\eeq
Suppose that
the spinor is subject to the $*$-gauge transformation
\beq
\psi(x)&\to& ^{\hat g}\psi(x)=U(x)*\psi(x),
\label{eqn:3-2}
\eeq
where $U(x)$ is assumed to be $*$-unitary:
\beq
U(x)*U^{\dag}(x)=
U^{\dag}(x)*U(x)=1.
\label{eqn:3-3}
\eeq
The $*$-gauge invariance of the Dirac action
leads to the replacement
of the partial derivative with the covariant one,
\beq
\partial_\mu\psi(x)&\to&
D_\mu\psi(x)=\partial_\mu\psi(x)
-ieA_\mu(x)*\psi(x),
\label{eqn:3-4}
\eeq
with the transformation law of the NC gauge field
\beq
A_\mu(x)\to
^{\hat g}\!\!A_\mu(x)=U(x)*A_\mu(x)*U^{\dag}(x)
+\frac ieU(x)*\partial_\mu U^{\dag}(x).
\label{eqn:3-5}
\eeq
This prescription gives the $*$-gauge-invariant Dirac action,
$S_{D_{*}}=S_{D_{0*}}+S_{I*}$, where
\beq
S_{I*}&=&
e\int\!d^4x
{\bar\psi}(x)*
\gamma^\mu A_\mu(x)*\psi(x)
\nn\\[2mm]
&=&e\int\!d^4x_1d^4x_2d^4x_3
K(x_1;x_2,x_3){\bar\psi}(x_1)
\gamma^\mu A_\mu(x_2)\psi(x_3).
\label{eqn:3-6}
\eeq
In the last expression we utilized the kernel\cite{7)} derived from
(\ref{eqn:1-2}) and (\ref{eqn:1-3}),
\begin{eqnarray}
K(x;x_1,x_2)&=&
\frac 1{(2\pi)^8}\int\!d^4p_1d^4p_2
e^{ip_1(x-x_1)+ip_2(x-x_2)}
e^{-\frac i2p_1\wedge p_2},
\label{eqn:3-7}
\end{eqnarray}
with $p_1\wedge p_2\equiv p_{1\mu}\theta^{\mu\nu}
p_{2\nu}$.
It can be shown that it is cyclic,
$K(x;x_1,x_2)=K(x_1;x_2,x)=K(x_2;x,x_1)$.
\\
\ind
At this stage we recall
the charge conjugation transformation of the spinor  $\psi$ 
and the (Abelian) gauge field $A_\mu$,
\beq
\!\!\!\!\!\!\!\!\!
\cC\psi(x)\cC^{-1}&=&C{\bar\psi}(x),\;\;\;
\cC{\bar\psi}(x)\cC^{-1}
=-\psi(x)C^{-1},\;\;\;
\cC A_\mu(x)\cC^{-1}=-A_\mu(x),
\label{eqn:3-8}
\eeq
respectively, where the charge conjugation matrix $C$ satisfies 
$C^{-1}\gamma^\mu C=-\gamma^{\mu T}$. 
If we resort to the sign change
(\ref{eqn:2-5}),
we find that
\beq
\cC S_{I*}\cC^{-1}
&=&-e\int\!d^4x_1d^4x_2d^4x_3
{\bar K}(x_1;x_2,x_3)\psi(x_1)
\gamma^{\mu T}A_\mu(x_2){\bar\psi}(x_3),
\label{eqn:3-9}
\eeq
where ${\bar K}(x_1;x_2,x_3)=K(x_1;x_2,x_3)|_{\theta\to -\theta}$.
As assumed in Ref. 19),
if the fields involved are classical,
commuting or anti-commuting at different $x$'s,
we may rewrite (\ref{eqn:3-9}) as
\beq
\cC S_{I*}\cC^{-1}
&=&e\int\!d^4x_1d^4x_2d^4x_3
{\bar K}(x_1;x_2,x_3){\bar\psi}(x_3)
\gamma^\mu A_\mu(x_2)\psi(x_1).
\label{eqn:3-10}
\eeq
It follows from (\ref{eqn:3-7})
that ${\bar K}(x_1;x_2,x_3)=K(x_1;x_3,x_2)$
which equals $K(x_3;x_2,x_1)$ by cyclicity.
Consequently, we finally have
\beq
\cC S_{I*}\cC^{-1}
&=&e\int\!\int\!d^4x_1d^4x_2d^4x_3
K(x_3;x_2,x_1){\bar\psi}(x_3)
\gamma^\mu A_\mu(x_2)\psi(x_1)
=S_{I*}.
\label{eqn:3-11}
\eeq
\ind
There is nothing to prevent us to
consider a possibility that
the fields neither commute nor anti-commute
at different $x$'s.
In such a case
$C$-invariant $*$-gauge interaction
should be the average,
\beq
S_I=\frac 12(S_{I*}+\cC S_{I*}\cC^{-1}).
\label{eqn:3-12}
\eeq
How to obtain (\ref{eqn:3-12}) via gauge principle
is solved as follows.
Let us write the free 
NC Dirac action $S_{D_0}$
as the average of (\ref{eqn:3-1}) and
\beq
S_{D_{0{\bar*}}}&=&
\int\!d^4x\psi(x){\bar*}
(i\gamma^{\mu T}\partial_\mu+m){\bar\psi}(x),
\label{eqn:3-13}
\eeq
that is, $S_{D_0}=\frac 12(S_{D_0*}
+S_{D_{0{\bar*}}})$.
Here we have introduced the anti-Moyal product, denoted ${\bar *}$,
\begin{eqnarray}
\varphi_1(x){\bar *}\varphi_2(x)
&\equiv&\varphi_2(x)*\varphi_1(x)
=
\varphi_1(x)e^{-\frac i2
\theta^{\mu\nu}{\mbox{\scriptsize$
{\overleftarrow{{\partial_\mu}}}$}}
{\mbox{\scriptsize$\overrightarrow{\partial_\nu}$}}}
\varphi_2(x).
\label{eqn:3-14}
\end{eqnarray}
The ${\bar *}$-product
satisfies the similar relations as those
obeyed by the $*$-product. In particular, the ${\bar *}$-product
is associative,
\beq
(\varphi_1(x){\bar *}\varphi_2(x)){\bar *}\varphi_2(x)
&=&\varphi_1(x){\bar *}(\varphi_2(x){\bar *}\varphi_2(x))
\equiv\varphi_1(x){\bar *}\varphi_2(x){\bar *}\varphi_2(x).
\label{eqn:3-15}
\end{eqnarray}
It is obvious that one can omit
both the symbols $*$ and ${\bar*}$ upon integration
under the same assumption,
\beq
\int\!d^4x\varphi_1(x)*\varphi_2(x)
&=&\int\!d^4x\varphi_1(x){\bar *}\varphi_2(x)
=\int\!d^4x\varphi_1(x)\varphi_2(x).
\label{eqn:3-16}
\end{eqnarray}
Since the definition (\ref{eqn:3-14})
works only for classical commuting functions,
we have to take into account
the extra minus sign appearing when exchanging the two spinors,
whence (\ref{eqn:3-13}) is equivalent to (\ref{eqn:3-1}).
\\
\ind
Next we introduce the covariant derivative
pertinent to the action (\ref{eqn:3-13}).
To this end we write the gauge transformations,
(\ref{eqn:3-2}) and (\ref{eqn:3-5}) in an `opposite'
but equivalent way,
\beq
\psi(x)&\to& ^{\hat g}\psi(x)
=\psi(x){\bar*}U(x),\nn\\[2mm]
A_\mu(x)&\to&
^{\hat g}\!\!A_\mu(x)=U^{\dag}(x){\bar*}
A_\mu(x){\bar*}U(x)
+\frac ie\partial_\mu U^{\dag}(x){\bar*}U(x).
\label{eqn:3-17}
\eeq
Note that $U(x)$ is also ${\bar *}$-unitary:
\beq
U(x){\bar*}U^{\dag}(x)=
U^{\dag}(x){\bar*}U(x)=1
\label{eqn:3-18}
\eeq
It is important to recognize that
the same $*$-gauge transformation
can also be written using the ${\bar*}$-product.
The $*$-gauge-invariant action based on (\ref{eqn:3-13})
is then obtained by
the replacement
\beq
\partial_\mu{\bar\psi}(x)&\to&
{\bar D}_\mu{\bar\psi}(x)=\partial_\mu{\bar\psi}(x)
+ieA_\mu(x){\bar*}{\bar\psi}(x).
\label{eqn:3-19}
\eeq
It is easy to prove that
\beq
{\bar D}_\mu{\bar\psi}(x)\to
U^{\dag}(x){\bar*}{\bar D}_\mu{\bar\psi}(x),
\label{eqn:3-20}
\eeq
under the ${\bar *}$-gauge transformation
(\ref{eqn:3-17}) with the unitarity (\ref{eqn:3-18}).
In total, the NC gauge-invariant Dirac action
is then given by
$S_D=
S_{D_0}+S_I$, where 
\beq
S_I&=&
\frac e2\int\!d^4x
[{\bar\psi}(x)*
\gamma^\mu A_\mu(x)*\psi(x)
-\psi(x){\bar*}\gamma^{\mu T}
A_\mu(x){\bar*}{\bar\psi}(x)]\nn\\[2mm]
&=&\frac e2\int\!d^4x_1d^4x_2d^4x_3
[K(x_1;x_2,x_3){\bar\psi}(x_1)
\gamma^\mu A_\mu(x_2)\psi(x_3)\nn\\[2mm]
&&\qquad\qquad\qquad-{\bar K}(x_1;x_2,x_3)\psi(x_1)\gamma^{\mu T}
A_\mu(x_2){\bar\psi}(x_3)].
\label{eqn:3-21}
\eeq
This is nothing but
(\ref{eqn:3-12}) noting (\ref{eqn:3-9}). 
Similarly, we have $P, T$ invariance. 
\\
\ind
Note that, under $CPT$, $K$ is changed into
${\bar K}$ because $CPT$ is anti-unitary
and the sign change (\ref{eqn:2-4})
is cancelled by (\ref{eqn:2-5}).
Remembering
$CPT$ transformation law,
\beq
&&\!\!\!\!\!\!\!\!\!\!\!\!\!\!\!\!\!\!\!\!\!\!\!\!\!\!\!
\Theta\psi(x)\Theta^{-1}=
-i\gamma_5\gamma_0^T{\bar\psi}(-x),\;\;\;
\Theta{\bar\psi}(x)\Theta^{-1}=
\psi(-x)i\gamma_5\gamma_0^T,\;\;\;
\Theta A_\mu(x)\Theta^{-1}=
-A_\mu(-x),
\label{eqn:3-22}
\eeq
with $\Theta\equiv\cC\cP\cT$,
we find that
\beq
\Theta S_{I*}\Theta^{-1}&=&
-e\int\!d^4x_1d^4x_2d^4x_3
{\bar K}(x_1;x_2,x_3)\psi(-x_1)
\gamma_0^T\gamma^{\mu *}\gamma_0^T
A_\mu(-x_2)*{\bar\psi}(-x_3)\nn\\[2mm]
&=&-e\int\!d^4x_1d^4x_2d^4x_3
{\bar K}(x_1;x_2,x_3)\psi(x_1)
\gamma^{\mu T}
A_\mu(x_2){\bar\psi}(x_3)
=S_{I{\bar*}}.
\label{eqn:3-23}
\eeq
For classical fields 
(commuting or anti-commuting at general $x$'s)
$S_{I*}=S_{I{\bar*}}$.
This proves $CPT$ invariance of QED$_*$
action in the fermion sector.
\\
\ind
The Maxwell sector is described by the action
\begin{eqnarray} 
S'_M&=&-\displaystyle{{1\over 4}}\int\!d^4x
F_{\mu\nu}(x)*F^{\mu\nu}(x),
\label{eqn:3-24}
\end{eqnarray}
where the Maxwell field strength
tensor is defined by
\begin{eqnarray} 
F_{\mu\nu}(x)&=&\partial_\mu A_\nu(x)
-\partial_\nu A_\mu(x)
      -ie[A_\mu(x),A_\nu(x)]_*,
\label{eqn:3-25}
\end{eqnarray}
with
\beq
&&[A_\mu(x),A_\nu(x)]_*\equiv
A_\mu(x)*A_\nu(x)-A_\nu(x)*A_\mu(x)\nn\\[2mm]
&&
=\int\!d^4x_1d^4x_2K(x;x_1,x_2)
[A_\mu(x_1)A_\nu(x_2)
-A_\nu(x_1)A_\mu(x_2)].
\label{eqn:3-26}
\end{eqnarray}
Under the charge conjugation
it goes over to
\beq
\cC[A_\mu(x),A_\nu(x)]_*\cC^{-1}
&=&\int\!d^4x_1d^4x_2{\bar K}(x;x_1,x_2)
\cC[A_\mu(x_1)A_\nu(x_2)
-A_\nu(x_1)A_\mu(x_2)]\cC^{-1}\nn\\[2mm]
&=&\int\!d^4x_1d^4x_2{\bar K}(x;x_1,x_2)
[A_\mu(x_1)A_\nu(x_2)
-A_\nu(x_1)A_\mu(x_2)]\nn\\[2mm]
&=&A_\mu(x){\bar*}A_\nu(x)-A_\nu(x){\bar*}A_\mu(x)
\equiv[A_\mu(x),A_\nu(x)]_{{\bar*}}.
\label{eqn:3-27}
\end{eqnarray}
Remember $K\to {\bar K}$ under $C$.
Consequently, the field strength
does not transform to itself
up to sign but is changed into 
\beq
\cC F_{\mu\nu}(x)\cC^{-1}&=&
-G_{\mu\nu}(x)\nn\\[2mm]
G_{\mu\nu}(x)&=&
\partial_\mu A_\nu(x)
-\partial_\nu A_\mu(x)
      +ie[A_\mu(x),A_\nu(x)]_{{\bar*}}.
\label{eqn:3-28}
\eeq
Using the last transformation of (\ref{eqn:3-17})
the field strength $G_{\mu\nu}(x)$
can be shown to be ${\bar*}$-gauge covariant,
\beq
G_{\mu\nu}(x)\to
{}^{\hat g}G_{\mu\nu}(x)
=U^{\dag}(x){\bar*}G_{\mu\nu}(x)
{\bar*}U(x).
\label{eqn:3-29}
\eeq
For classical fields we can commute
two field variables
in (\ref{eqn:3-27}) to find 
$F_{\mu\nu}=G_{\mu\nu}$ as observed in Ref. 19).
For non-classical fields which are noncommuting at general
$x$'s we have $F_{\mu\nu}\ne G_{\mu\nu}$.
Thus $C$-invariant Maxwell$_*$ action also becomes
the average
\begin{eqnarray} 
S_M&=&-\displaystyle{{1\over 8}}
\int\!d^4x
[F_{\mu\nu}(x)*F^{\mu\nu}(x)
+G_{\mu\nu}(x){\bar*}G{}^{\mu\nu}(x)].
\label{eqn:3-30}
\end{eqnarray}
Hence, we have shown that
\beq
\cC S_{QED}\cC^{-1}=S_{QED},\;\;\;
S_{QED}=S_D+S_M.
\label{eqn:3-31}
\eeq
This proves $C$-invariance of QED$_*$.
\\
\ind
As a final example we consider $CPT$ in the Maxwell sector.
Using the last equation of (\ref{eqn:3-22}) and $K\to {\bar K}$
under $CPT$, we have
\beq
\Theta F_{\mu\nu}(x)\Theta^{-1}&=&
G_{\mu\nu}(-x),
\label{eqn:3-32}
\eeq
where $G_{\mu\nu}(x)$ is defined by (\ref{eqn:3-28}).
Hence the Maxwell$_*$ action $S_M$ is $CPT$-invariant.
In conjunction with the fermion sector we
have proved
\beq
\Theta S_{QED}\Theta^{-1}=S_{QED},\;\;\;
S_{QED}=S_D+S_M.
\label{eqn:3-33}
\eeq
Thus QED$_*$ conserves $CPT$
despite of the Lorentz violation.\cite{19),20),21)}
The proof hinges upon the transformations
(\ref{eqn:2-3}) and
(\ref{eqn:2-4}) (and the sign change (\ref{eqn:2-5}))
which, however, can not be derived
within the framework of the $\theta$-algebra.
In this sense the discrete symmetries in
QED$_*$ do not match our symmetry criterion in the section 1.
We amend these points in the following sections.
\section{Discrete symmetries of the DFR algebra}
Although the $\theta$-algebra (\ref{eqn:1-1})
is not
invariant under $P$, and $T$, separately,
unless the transformations
(\ref{eqn:2-3}) and (\ref{eqn:2-4})
are taken into account simultaneously,
the DFR algebra (\ref{eqn:1-7}) has the discrete symmetries
if we extend the action of the operators,
$\cP, \cT, \cC$, to the operators
defined on the DFR algebra (\ref{eqn:1-7}) such that
\begin{eqnarray}
\cP\,\htheta^{\mu\nu}\cP^{-1}&=&\htheta_{\mu\nu},\;\;\;
\cT\,\htheta^{\mu\nu}\cT^{-1}=-\htheta_{\mu\nu},\nn\\[2mm]
\cC\;\htheta^{\mu\nu}\cC^{-1}&=&\htheta^{\mu\nu},\;\;\;
\Theta\;\htheta^{\mu\nu}\Theta^{-1}=-\htheta^{\mu\nu}.
\label{eqn:4-1}
\end{eqnarray}
Considering
the $\theta$-space spanned by the eigenstates
of the operator $\htheta^{\mu\nu}$ as in Ref. 13), (\ref{eqn:4-1})
is equivalent to (\ref{eqn:2-3}) and (\ref{eqn:2-4}).
We again emphasize that the non-trivial transformations 
(\ref{eqn:2-3}) and (\ref{eqn:2-4})
of the components of $\theta^{\mu\nu}$
cannot be obtained within the framework of the $\theta$-algebra.
It is associated with the facts that
the discrete symmetry operators can be extended to
act on the operators defined on the DFR algebra
and that $\theta^{\mu\nu}$ is an eigenvalue of the
operator $\htheta^{\mu\nu}$.
\section{Discrete symmetries in Lorentz-invariant QED$_*$}
The purpose of this section
is to prove the discrete symmetries
of Lorentz-invariant QED$_*$.
The proof differs from that of QED$_*$
in that both $x$ and $\theta$
become the integration variables in the action (\ref{eqn:1-9}).
\\
\ind
Let us start from the Lorentz-invariant free Dirac action
\beq
\hS_{D_{0*}}&=&
\int\!d^4xd^6\theta W(\theta){\bar\psi}(x,\theta)*
(i\gamma^\mu\partial_\mu-m)\psi(x,\theta)\nn\\[2mm]
&=&\int\!d^6\theta W(\theta)\int\!d^4x_1d^4x_2d^4x_3
K(x_1;x_2,x_3){\bar\psi}(x_1,\theta)
(i\gamma^\mu\partial_{2\mu}-m)\psi(x_2,\theta).
\label{eqn:5-1}
\eeq
An explicit expression
in terms of the kernel
can be derived from (\ref{eqn:1-10}).
Now suppose that
the spinor is subject to the $*$-gauge transformation
\beq
\psi(x,\theta)&\to& ^{\hat g}\psi(x,\theta)
=U(x,\theta)*\psi(x,\theta),\nn\\[2mm]
{\bar\psi}(x,\theta)&\to& ^{\hat g}{\bar\psi}(x,\theta)
={\bar\psi}(x,\theta)*U^{\dag}(x,\theta),
\label{eqn:5-2}
\eeq
where $U(x,\theta)$ is assumed to be $*$-unitary:
\beq
U(x,\theta)*U^{\dag}(x,\theta)=
U^{\dag}(x,\theta)*U(x,\theta)=1.
\label{eqn:5-3}
\eeq
The $*$-gauge invariance requires
the replacement
\beq
\partial_\mu\psi(x,\theta)&\to&
D_\mu\psi(x,\theta)=\partial_\mu\psi(x,\theta)
-ieA_\mu(x,\theta)*\psi(x,\theta),
\label{eqn:5-4}
\eeq
with the NC gauge field transforming like
\beq
A_\mu(x,\theta)\to
^{\hat g}\!\!A_\mu(x,\theta)&=&U(x,\theta)*A_\mu(x,\theta)*
U^{\dag}(x,\theta)
+\frac ieU(x,\theta)*\partial_\mu U^{\dag}(x,\theta).
\label{eqn:5-5}
\eeq
This prescription gives the $*$-gauge invariant
Dirac action, $\hS_{D*}=\hS_{D_{0*}}+\hS_{I*}$, 
where
\beq
\hS_{I*}&=&
e\int\!d^4xd^6\theta W(\theta)
{\bar\psi}(x,\theta)*
\gamma^\mu A_\mu(x,\theta)*\psi(x,\theta)\nn\\[2mm]
&=&e\int\!d^6\theta W(\theta)\int\!d^4x_1d^4x_2d^4x_3
K(x_1;x_2,x_3){\bar\psi}(x_1,\theta)
\gamma^\mu A_\mu(x_2,\theta)\psi(x_3,\theta).
\label{eqn:5-6}
\eeq
To investigate $C$-transformation property of this action
we recall
$C$-transformation\cite{12)} of the fields in Lorentz-invariant
spinor QED$_*$,
\beq
\cC\psi(x,\theta)\cC^{-1}
&=&C{\bar\psi}(x,\theta),\nn\\[2mm]
\cC{\bar\psi}(x,\theta)\cC^{-1}
&=&-\psi(x,\theta)C^{-1},\nn\\[2mm]
\cC A_\mu(x,\theta)\cC^{-1}
&=&-A_\mu(x,\theta).
\label{eqn:5-7}
\eeq
Since the sign change (\ref{eqn:2-5})
is relevant only for the kernel,
$C$ transformation (\ref{eqn:5-7}) keeps the sign of 
the argument $\theta$ of fields
as indicated by (\ref{eqn:2-3}). 
We have using (\ref{eqn:2-5})
\beq
\cC\hS_{I*}\cC^{-1}&=&
-e\int\!d^4xd^6\theta W(\theta)
\psi(x,\theta){\bar*}
\gamma^{\mu T}A_\mu(x,\theta){\bar*}{\bar\psi}(x,\theta)\nn\\[2mm]
&=&-e\int\!d^6\theta W(\theta)\int\!d^4x_1d^4x_2d^4x_3
{\bar K}(x_1;x_2,x_3)\psi(x_1,\theta)
\gamma^{\mu T}A_\mu(x_2,\theta){\bar\psi}(x_3,\theta).
\label{eqn:5-8}
\eeq
For classical fields this is further rearranged into
\beq
\cC\hS_{I*}\cC^{-1}
&=&e\int\!d^6\theta W(\theta)\int\!d^4x_1d^4x_2d^4x_3
{\bar K}(x_1;x_2,x_3){\bar\psi}(x_3,\theta)
\gamma^\mu A_\mu(x_2,\theta)\psi(x_1,\theta).
\label{eqn:5-9}
\eeq
Noting that ${\bar K}(x_1;x_2,x_3)=
K(x_1;x_3,x_2)=K(x_3;x_2,x_1)$ by cyclicity,
the right-hand side equals $\hS_{I*}$:
\beq
\cC\hS_{I*}\cC^{-1}
&=&\hS_{I*}.
\label{eqn:5-10}
\eeq
\ind
For non-classical field we can not make the rearrangement
from (\ref{eqn:5-8}) to (\ref{eqn:5-9}) and are unable
to arrive at the result (\ref{eqn:5-10}).
Hence $C$-invariant NC Dirac action in a Lorentz-invariant version
is also the average
\beq
\hS_D=\frac 12[(\hS_{D_{0*}}+\hS_{I*})+\cC(\hS_{_{D0*}}+\hS_{I*})\cC^{-1}].
\label{eqn:5-11}
\eeq
For completeness we explicitly write the third term,
\beq
\hS_{D_{0{\bar*}}}&\equiv&\cC\hS_{D0*}\cC^{-1}
=
\int\!d^4xd^6\theta W(\theta)
\psi(x,\theta){\bar*}
(i\gamma^{\mu T}\partial_\mu+m){\bar\psi}(x,\theta)\nn\\[2mm]
&=&\int\!d^6\theta W(\theta)\int\!d^4x_1d^4x_2d^4x_3
{\bar K}(x_1;x_2,x_3)\psi(x_1,\theta)
(i\gamma^{\mu T}\partial_{2\mu}+m){\bar\psi}(x_2,\theta).
\label{eqn:5-12}
\eeq
The reason why the ${\bar*}$-product
appears here is the same as in (\ref{eqn:3-13}).
Following the procedure in the section 3,
\footnote{We may add the argument $\theta$
to the functions in (\ref{eqn:3-14}), (\ref{eqn:3-15}), and (\ref{eqn:3-16}) 
with appropriate care.}
we write the $*$-gauge transformations (\ref{eqn:5-2})
and (\ref{eqn:5-5})
in an `opposite' but equivalent way
\beq
\psi(x,\theta)&\to& ^{\hat g}\psi(x,\theta)
=\psi(x,\theta){\bar*}U(x,\theta),\nn\\[2mm]
{\bar\psi}(x,\theta)&\to& ^{\hat g}{\bar\psi}(x,\theta)
=U^{\dag}(x,\theta){\bar*}{\bar\psi}(x,\theta),\nn\\[2mm]
A_\mu(x,\theta)&\to&
^{\hat g}\!\!A_\mu(x,\theta)=U^{\dag}(x,\theta){\bar*}
A_\mu(x,\theta){\bar*}U(x,\theta)
+\frac ie\partial_\mu U^{\dag}(x,\theta){\bar*}U(x,\theta).
\label{eqn:5-13}
\eeq
where $U(x,\theta)$ is also ${\bar *}$-unitary:
\beq
U(x,\theta){\bar*}U^{\dag}(x,\theta)=
U^{\dag}(x,\theta){\bar*}U(x,\theta)=1.
\label{eqn:5-14}
\eeq
The $*$-gauge-invariant action is then obtained by
the replacement
\beq
\partial_\mu{\bar\psi}(x,\theta)&\to&
{\bar D}_\mu{\bar\psi}(x,\theta)=\partial_\mu{\bar\psi}(x,\theta)
+ieA_\mu(x,\theta){\bar*}{\bar\psi}(x,\theta),
\label{eqn:5-15}
\eeq
in (\ref{eqn:5-12}). The covariant derivative ${\bar D}_\mu$
satisfies
\beq
{\bar D}_\mu{\bar\psi}(x,\theta)\to
U^{\dag}(x,\theta){\bar*}{\bar D}_\mu{\bar\psi}(x,\theta),
\label{eqn:5-16}
\eeq
under the ${\bar *}$-gauge transformation
(\ref{eqn:5-13}).
The result yields $\hS_{D{\bar*}}
=\hS_{D_0{\bar*}}+\hS_{I{\bar*}}$, where
\beq
\hS_{I{\bar*}}&=&
-e\int\!d^4xd^6\theta W(\theta)
\psi(x,\theta){\bar*}\gamma^{\mu T}
A_\mu(x,\theta){\bar*}{\bar\psi}(x,\theta)\nn\\[2mm]
&=&-e\int\!d^6\theta W(\theta)\int\!d^4x_1d^4x_2d^4x_3
{\bar K}(x_1;x_2,x_3)\psi(x_1,\theta)\gamma^{\mu T}
A_\mu(x_2,\theta){\bar\psi}(x_3,\theta).
\label{eqn:5-17}
\eeq
This is nothing but (\ref{eqn:5-8}).
Hence if we put $\hS_D=\frac 12(\hS_{D*}+\hS_{D{\bar*}})$,
it is now clear that
\beq
\cC\hS_D\cC^{-1}=\hS_D,
\label{eqn:5-18}
\eeq
provided
$*$-product$\leftrightarrow{\bar*}$-product, i.e.,
$K\leftrightarrow {\bar K}$
under $C$.
\\
\ind
Before proceeding further we ask ourselves why the sign change
(\ref{eqn:2-5}) is necessary in proving
$C$ invariance of the action $\hS_D$.
Although the free actions (\ref{eqn:5-1}) and (\ref{eqn:5-12})
are the same, the nonlocal Lagrangians corresponding to them
are different,
\beq
\hL_{D_{0*}}&=&
{\bar\psi}(x,\theta)*
(i\gamma^\mu\partial_\mu-m)\psi(x,\theta)\nn\\[2mm]
&=&\int\!d^4x_1d^4x_2K(x;x_1,x_2)
{\bar\psi}(x_1,\theta)(i\gamma^\mu\partial_{2\mu}-m)\psi(x_2,\theta)
,\nn\\[2mm]
\hL_{D_{0{\bar*}}}&=&
\psi(x,\theta){\bar*}
(i\gamma^{\mu T}\partial_\mu+m){\bar\psi}(x,\theta)\nn\\[2mm]
&=&\int\!d^4x_1d^4x_2{\bar K}(x;x_1,x_2)
\psi(x_1,\theta)(i\gamma^{\mu T}\partial_{2\mu}+m){\bar\psi}(x_2,\theta).
\label{eqn:5-19}
\eeq
We are going to compare these nonlocal Lagrangians 
in connection with $C$.
This is in the same spirit as in the commutative theory
where
the Dirac Lagrangian omitting $*$ in (\ref{eqn:3-1})
is $C$-invariant only up to a total divergence
and
$C$-invariant free Dirac Lagrangian
is the average,
\footnote{This is usually understood when writing the
Dirac Lagrangian without the antisymmetrization. 
We would like to point out, however,
that a similar `averaging' involves a nontrivial
prescription in the NC setting.}
\beq
\cL_{D_0}&=&
\frac 12[{\bar\psi}(x)
(i\gamma^\mu\partial_\mu-m)\psi(x)
+\psi(x)
(i\gamma^{\mu T}\partial_\mu+m){\bar\psi}(x)].
\label{eqn:5-20}
\eeq
Our procedure to obtain $C$-invariant free Dirac Lagrangian
in the NC setting goes through in a similar way.
If we transform
(\ref{eqn:5-19}) without touching the kernel,
we obtain an unpleasant result that
$\cC\hL_{D_{0*}}\cC^{-1}$ is neither equal to $\hL_{D_{0*}}$ 
itself nor transformed to
$\hL_{D_{0{\bar*}}}$ and similarly for $\cC\hL_{D_{0{\bar*}}}\cC^{-1}$.
Instead we require as in the commutative case that
$\hL_{D_{0*}}$ be transformed to $\hL_{D_{0{\bar*}}}$ and vise versa
under $C$.
This requirement is satisfied by assuming that
the kernel $K(x;x_1,x_2)$ is changed into
the kernel ${\bar K}(x;x_1,x_2)$ and vise versa under $C$.
This is seen as follows.
\beq
\cC\hL_{D_{0*}}\cC^{-1}
&=&\int\!d^4x_1d^4x_2{\bar K}(x;x_1,x_2)
[-\psi(x_1,\theta)C^{-1}(i\gamma^\mu\partial_{2\mu}-m)C{\bar\psi}(x_2,\theta)]
\nn\\[2mm]
&=&\int\!d^4x_1d^4x_2{\bar K}(x;x_1,x_2)
\psi(x_1,\theta)(i\gamma^{\mu T}\partial_{2\mu}+m)
{\bar\psi}(x_2,\theta)
=\hL_{D_{0{\bar*}}},\nn\\[2mm]
\cC\hL_{D_{0{\bar*}}}\cC^{-1}
&=&\int\!d^4x_1d^4x_2K(x;x_1,x_2)
[-C{\bar\psi}(x_1,\theta)(i\gamma^{\mu T}\partial_{2\mu}+m)\psi(x_2,\theta)C^{-1}]\nn\\[2mm]
&=&\int\!d^4x_1d^4x_2K(x;x_1,x_2)
{\bar\psi}(x_1,\theta)(i\gamma^\mu\partial_{2\mu}-m)\psi(x_2,\theta)
=\hL_{D_{0*}}.
\label{eqn:5-21}
\eeq
The exchange $K\leftrightarrow {\bar K}$
under $C$ corresponds to the sign change (\ref{eqn:2-5}),
which resembles an anti-unitary nature,
$i\to -i$ under $T$ since we keep the sign
of the argument $\theta$ in the field variables.
\footnote{This is reminiscent of the anti-unitary character
of the charge conjugation in the one-particle theory.}
\\
\ind
The Maxwell$_*$ action is constructed using the field strength
tensor
\begin{eqnarray} 
F_{\mu\nu}(x,\theta)&=&\partial_\mu A_\nu(x,\theta)
-\partial_\nu A_\mu(x,\theta)
      -ie[A_\mu(x,\theta),A_\nu(x,\theta)]_*,
\label{eqn:5-22}
\end{eqnarray}
where 
\beq
[A_\mu(x,\theta),A_\nu(x,\theta)]_*\equiv
A_\mu(x,\theta)*A_\nu(x,\theta)-A_\nu(x,\theta)*A_\mu(x,\theta)
\label{eqn:5-23}
\end{eqnarray}
is the Moyal bracket, so that
\begin{eqnarray} 
{\hat S}'_M&=&-\displaystyle{{1\over 4}}\int\!d^4xd^6\theta W(\theta)
F_{\mu\nu}(x,\theta)*F^{\mu\nu}(x,\theta).
\label{eqn:5-24}
\end{eqnarray}
The non-linear term in the
field strength may be written as
\beq
\!\!\!\!\!\!\!\!\!
[A_\mu(x,\theta),A_\nu(x,\theta)]_*
&=&\int\!d^4x_1d^4x_2K(x;x_1,x_2)
[A_\mu(x_1,\theta)A_\nu(x_2,\theta)
-A_\nu(x_1,\theta)A_\mu(x_2,\theta)].
\label{eqn:5-25}
\end{eqnarray}
Under the charge conjugation
it goes over to
\beq
\cC[A_\mu(x,\theta),A_\nu(x,\theta)]_*\cC^{-1}
&=&\int\!d^4x_1d^4x_2{\bar K}(x;x_1,x_2)
\cC[A_\mu(x_1,\theta)A_\nu(x_2,\theta)
-A_\nu(x_1,\theta)A_\mu(x_2,\theta)]\cC^{-1}\nn\\[2mm]
&=&\int\!d^4x_1d^4x_2{\bar K}(x;x_1,x_2)
[A_\mu(x_1,\theta)A_\nu(x_2,\theta)
-A_\nu(x_1,\theta)A_\mu(x_2,\theta)]\nn\\[2mm]
&=&A_\mu(x,\theta){\bar*}A_\nu(x,\theta)-A_\nu(x,\theta)
{\bar*}A_\mu(x,\theta)
\equiv[A_\mu(x,\theta),A_\nu(x,\theta)]_{{\bar*}}.
\label{eqn:5-26}
\end{eqnarray}
Remember $K\to {\bar K}$ under $C$.
Consequently, as in the section 3,
the field strength
does not transform to itself
up to sign but is changed into 
\beq
\cC F_{\mu\nu}(x,\theta)\cC^{-1}&=&
-G_{\mu\nu}(x,\theta)\nn\\[2mm]
G_{\mu\nu}(x,\theta)&=&
\partial_\mu A_\nu(x,\theta)
-\partial_\nu A_\mu(x,\theta)
      +ie[A_\mu(x,\theta),A_\nu(x,\theta)]_{{\bar*}}.
\label{eqn:5-27}
\eeq
Using the last transformation of (\ref{eqn:5-13})
the field strength $G_{\mu\nu}(x,\theta)$
can be shown to be ${\bar*}$-gauge covariant,
\beq
G_{\mu\nu}(x,\theta)\to
{}^{\hat g}G_{\mu\nu}(x,\theta)
=U^{\dag}(x,\theta){\bar*}G_{\mu\nu}(x,\theta)
{\bar*}U(x,\theta).
\label{eqn:5-28}
\eeq
For classical fields we can commute
two field variables
in (\ref{eqn:5-26}) to find 
$F_{\mu\nu}=G_{\mu\nu}$ as observed in Ref. 12).
For non-classical fields which are noncommuting at general
$x$'s we have $F_{\mu\nu}\ne G_{\mu\nu}$.
Thus $C$-invariant Maxwell$_*$ action also becomes
the average
\begin{eqnarray} 
\!\!\!\!\!\!\!\!\!
{\hat S}_M&=&-\displaystyle{{1\over 8}}
\int\!d^4xd^6\theta W(\theta)
[F_{\mu\nu}(x,\theta)*F^{\mu\nu}(x,\theta)
+G_{\mu\nu}(x,\theta){\bar*}G{}^{\mu\nu}(x,\theta)].
\label{eqn:5-29}
\end{eqnarray}
Hence, we have shown that
\beq
\cC {\hat S}_{QED}\cC^{-1}={\hat S}_{QED},\;\;\;
{\hat S}_{QED}={\hat S}_D+{\hat S}_M.
\label{eqn:5-30}
\eeq
This proves $C$-invariance of Lorentz-invariant QED$_*$.
\\
\ind
Next we turn to $P, T$ and $CPT$.
We assume that
\beq
\cP\psi(x,\theta)\cP^{-1}&=&
\gamma^0\psi(x_P,\theta_P),\;\;\;
\cP{\bar\psi}(x,\theta)\cP^{-1}=
{\bar\psi}(x_P,\theta_P)\gamma^0,\;\;\;
\cP A_\mu(x,\theta)\cP^{-1}=
A^\mu(x_P,\theta_P),\nn\\[2mm]
\cT\psi(x,\theta)\cT^{-1}&=&
R\psi(x_T,\theta_T),\;\;\;
\cT\psi^{\dag}(x,\theta)\cT^{-1}=
\psi^{\dag}(x_T,\theta_T)R^{-1},\;\;\;
\cT A_\mu(x,\theta)\cT^{-1}=
A^\mu(x_T,\theta_T),\nn\\
&&
\label{eqn:5-31}
\eeq
with $
\theta_P\equiv(\theta^{\mu\nu})_P=\theta_{\mu\nu},
\theta_T\equiv(\theta^{\mu\nu})_T=-\theta_{\mu\nu}$
and $R^{-1}\gamma^\mu{}^* R=\gamma_\mu$ with $R=i\gamma_5C$.
The Dirac action $\hS_D=
\frac 12(\hS_{D_{*}}+\hS_{D_{\bar*}})$
and the Maxwell$_*$ action $\hS_M$ 
are invariant under $P$ and $T$.
\footnote{We assume that $W(\theta)$
is a function of the invariant
$\theta^{\mu\nu}\theta_{\mu\nu}$ only
so that $W(\theta_P)=W(\theta_T)=W(\theta)$.}
Hence $P$ and $T$ are separately conserved
in Lorentz-invariant QED$_*$. 
\\
\ind
The $CPT$ transformation is determined from 
(\ref{eqn:5-7}) and (\ref{eqn:5-31}) to be
\beq
\Theta\psi(x,\theta)\Theta^{-1}&=&
-i\gamma_5\gamma_0^T{\bar\psi}(-x,-\theta),\nn\\[2mm]
\Theta{\bar\psi}(x,\theta)\Theta^{-1}&=&
\psi(-x,-\theta)i\gamma_5\gamma_0^T,\nn\\[2mm]
\Theta A_\mu(x,\theta)\Theta^{-1}&=&
-A_\mu(-x,-\theta).
\label{eqn:5-32}
\eeq
We note the difference
from the prescription used in (\ref{eqn:3-23})
in that now $K$ is unchanged under $CPT$
because of the simultaneous change $i\to -i$
and (\ref{eqn:2-5}). The $CPT$
transformation (\ref{eqn:2-4})
is already taken into account in (\ref{eqn:5-32}).
The calculation will be reported only 
in the fermion sector,
\beq
\Theta\hS_{D_*}\Theta^{-1}&=&
-e\int\!d^6\theta W(\theta)
\int\!d^4x_1d^4x_2d^4x_3
K(x_1;x_2,x_3)
\psi(-x_1,-\theta)
\gamma_0^T\gamma^{\mu *}\gamma_0^T
A_\mu(-x_2,-\theta){\bar\psi}(-x_3,-\theta)\nn\\[2mm]
&=&
-e\int\!d^6\theta W(\theta)
\int\!d^4x_1d^4x_2d^4x_3
{\bar K}(x_1;x_2,x_3)
\psi(x_1,\theta)
\gamma^{\mu T}
A_\mu(x_2,\theta){\bar\psi}(x_3,\theta)
=\hS_{D{\bar*}},
\label{eqn:5-33}
\eeq
where we have used the fact that
the weight function is even, $W(-\theta)=W(\theta)$.
Consequently, we have shown that
\beq
\Theta\hS_{\rm QED}\Theta^{-1}
=\hS_{\rm QED}, \;\;\;\hS_{\rm QED}=\hS_D+\hS_M.
\label{eqn:5-34}
\eeq
Lorentz-invariant QED$_*$ is $CPT$-invariant
in accord with our definition of the symmetry described 
in the section 1.
\section{Conclusions}
We have discussed the discrete symmetries in Lorentz-invariant
non-commutative field theory based on the DFR algebra.
Since
anti-particles are an outcome from the marriage
of relativity and quantum mechanics,
the concept of the
charge conjugation 
can only be defined in relativistic quantum field theory
even if the continuum space-time
is modified to the non-commutative one.
In this respect we differ from the previous
discussions\cite{19),20),24)} on the same subject.
\\
\ind
The spinor $\psi$ in the previous section
is nothing but
$\psi_1$ in Ref. 12) with $e\to 2e$.
It was shown there that there are only eight spinors
allowed in Lorentz-invariant QED$_*$
compatible with $*$-gauge transformations.
Three of them
couple to $A_\mu(x,\theta)$
and the other three to
$A'_\mu(x,\theta)=-A_\mu(x,-\theta)$. Both obey the charge 
quantization condition.\cite{25)}
The neutral spinor among them
may represent the neutrinos.
There are two more spinors either of which may
be identified with the observed charged leptons.
\\
\ind
Our discussion is thus restricted to a particular sector
of Lorentz-invariant QED$_*$.
Nonetheless, it is straightforward to extend it to
generic QFT$_*$.
\\
\ind
At present stage we are unable to find a consistent way of quantization
except when the field is independent of $\theta$.
This is an important problem in our formalism
and we shall come back to it in later communication.
For phenomenological purpose\cite{11),12),13)} we may use 
the $\theta$-expansion, that is,
a field redefinition which
expresses any field with the
`internal coordinates' $\theta$ in terms of
those without the
`internal coordinates'. For a generic field $\varphi(x,\theta)$
occurring in the $*$-gauge theory
this is accomplished by assuming
$$
\varphi(x,\theta)=
\varphi^{(0)}(x)+\varphi^{(1)}(x)
+\varphi^{(2)}(x)+\cdots,
$$
where $\varphi^{(n)}(x)$ is of order $n$ in
$\theta, n=0,1,2,\cdots,$ and is
a function of the lowest-order fields 
and their derivatives.
Consequently, it is only necessary to second quantize
the local field $\varphi(x)\equiv\varphi^{(0)}(x)$.
One can do this using the Seiberg-Witten map.\cite{2)}
%
\section*{Acknowledgements}
The author is grateful to
H. Kase, Y. Okumura and E. Umezawa
for useful discussions.

\end{document}